\documentclass[twocolumn,showpacs,preprintnumbers,amsmath,amssymb]{revtex4}
\usepackage{graphicx}
\usepackage{amsmath}
\def\apj #1 #2 #3 {#1, ApJ, {\bf #2}, #3}
\def\apjl #1 #2 #3 {#1, ApJ, {\bf #2}, L#3}
\def\apjs #1 #2 #3 {#1, ApJS, {\bf #2}, #3}
\def\aap  #1 #2 #3 {#1, A\&A, {\bf #2}, #3}
\def\mnras #1 #2 #3 {#1, MNRAS, {\bf #2}, #3}
\def\pra #1 #2 #3 {#1, Phys.~Rev.~A., {\bf #2}, #3}
\def\prb #1 #2 #3 {#1, Phys.~Rev.~B., {\bf #2}, #3}
\def\prc #1 #2 #3 {#1, Phys.~Rev.~C., {\bf #2}, #3}
\def\prd #1 #2 #3 {#1, Phys.~Rev.~D., {\bf #2}, #3}
\def\pre #1 #2 #3 {#1, Phys.~Rev.~E., {\bf #2}, #3}
\def\prl #1 #2 #3 {#1, Phys.~Rev.~Lett., {\bf #2}, #3}
\def\plb #1 #2 #3 {#1, Phys.~Lett.~B., {\bf #2}, #3}
\def\science #1 #2 #3 {#1, Science., {\bf #2}, #3}
\def\nature #1 #2 #3 {#1, Nature., {\bf #2}, #3}
\def\nphysa #1 #2 #3 {#1, Nucl.~Phys.~A., {\bf #2}, #3}
\def\nphysb #1 #2 #3 {#1, Nucl.~Phys.~B., {\bf #2}, #3}
\def\nphysbs #1 #2 #3 {#1, Nucl.~Phys.~B.~Suppl., {\bf #2}, #3}

\def\h#1{\hbox{${}^{#1}$H}}
\def\he#1{\hbox{${}^{#1}$He}}
\def\li#1{\hbox{${}^{#1}$Li}}
\def\be#1{\hbox{${}^{#1}$Be}}

\def\dovh{\hbox{D/H}}

\def\yp{\hbox{\hbox{$Y_p$}}}

\def\h502{\hbox{$ h^{2}_{50}$}}

\def\la{\mathrel{\mathpalette\fun <}}
\def\ga{\mathrel{\mathpalette\fun >}}
\def\fun#1#2{\lower3.6pt\vbox{\baselineskip0pt\lineskip.9pt
  \ialign{$\mathsurround=0pt#1\hfil##\hfil$\crcr#2\crcr\sim\crcr}}}
\begin{document}
\draft
\title{Cosmological Constraints on Newton's  Constant}
\author{ Ken-ichi Umezu$^{1,2}$, Kiyotomo Ichiki$^{2,3}$, Masanobu Yahiro$^{4}$}
\address{
$^1$Department of Astronomical Science, the Graduate University for
Advanced Studies, 2-21-1, Osawa, Mitaka, Tokyo 181-8588, Japan
}
\address{
$^2$National Astronomical Observatory, 2-21-1, Osawa, Mitaka, Tokyo
181-8588, Japan
}
\address{
$^3$University of Tokyo, Department of Astronomy, 7-3-1 Hongo,
Bunkyo-ku, Tokyo 113-0033, Japan
}
\address{
$^4$Department of Physics, Kyushu University, Hakozaki, Higashi-ku,
Fukuoka 812-8518, Japan
}
\date{\today}
\begin{abstract}
 We present cosmological constraints on deviations of Newton's
 constant at large scales, analyzing latest cosmic microwave
 background (CMB) anisotropies and primordial abundances of light elements
 synthesized by big bang nucleosynthesis (BBN).
 BBN limits the possible deviation at typical scales of 
BBN epoch, say at $10^8 \sim 10^{12}$ m, to lie 
between $-5\%$ and $+1\%$ of the experimental value, and 
CMB restricts the deviation at larger scales 
$10^2 \sim 10^{9}$ pc 
to be between $-26\%$ and $+66\%$ 
at the $2\sigma$ confidence level.
The cosmological constraints are compared with the astronomical one from 
the evolution of isochrone of globular clusters.

 \end{abstract}
\maketitle
%
%
%

Newton's law of gravitation has been extensively tested and verified 
in three length scales: the laboratory scales $r \la 1$ m
\cite{Long:2003dx}, the geophysical scales $r \approx 100$ m
\cite{Baldi:2001rh}, and the astronomical scales $r \approx 10^8$ m 
\cite{2002nmgm.meet.1797W}. 
Such measurements nicely agree with the inverse square law within their
experimental or observational uncertainty 
\cite{1999snng.book.....F,Adelberger:2003zx}.
In particular, the first two measurements at the laboratory and geophysical 
scales succeeded also in determining the experimental value $G_{\rm N}$ 
of Newton's constant, and the value determined at such terrestrial
scales is applied for all phenomena from Planck scale to cosmological
scale.

The astronomical measurements \cite{2002nmgm.meet.1797W}, mainly through 
planetary and satellite orbits, yield a strong constraint on 
the deviation from the inverse square law. 
However, it should be noted that 
the measurements can not 
give any information about the value of Newton's constant $G$ itself
without evaluating masses $M$ of interacting bodies, 
since constraint is possible only on $GM$.
Therefore, the  measurements can not exclude the possibility of different
value of $G$ at astronomical and cosmological scales, 
if $G$ is almost constant at limited scales relevant to the measurements. 
In particular, we have only a poor
knowledge at scales larger than the solar system, say $r \ga 1$ pc
$\approx 3\times10^{16}$ m.
\cite{Adelberger:2003zx}. Interesting trials to solve this problem were
recently reported \cite{Shirata:2005yr,Sealfon:2004gz}, in which 
the deviation of $G$ at Mpc scales is 
restricted by the power of the clustering of galaxies .

The possibility that Newton's constant at laboratory scale, $G_{\rm N}$,
is different from that at very large scales, $G_{\infty}$, arises in
many context. Historically, studies toward the problem of unifying
gravity with the other fundamental forces suggested a departure from
Newtonian gravity in the range $10-100$ m.
\cite{Wagoner:1970vr}.
It is often assumed that such a correction can be represented by
the addition of Yukawa term to the conventional gravitational potential:
 $V = - \frac{G(r) M}{r}$ for 
 $G(r)=G_\infty\left(1+\alpha e^{-r/\lambda}\right),$
where 
$\alpha$ is the relative weight of the non-Newtonian term.
In this expression, at  cosmological distances $r$ satisfying 
$r \gg \lambda$, the exponential term vanishes, so that $G(r)=G_\infty$. 
On the other hand,  for $r$ of 
experimental scales which satisfies  $r \ll \lambda$, 
the exponential becomes unity and $G(r)$ recovers $G_{\rm N}$, that is,  
$ G_{\rm N} = G_\infty\left(1+\alpha\right)~.$

Recently several types of higher-dimensional theories of gravity,
motivated by superstring, have been proposed and many researchers pay 
great attention to the extra dimension scenario. As a characteristic
feature, all the theories lead to  deviations from the conventional
Newton's law \cite{Garriga:1999yh,Ghoroku:2003bs}, since the theories 
allow graviton to propagate in higher-dimensional spacetime. 
Among them, an interesting idea was proposed by Dvali, Gabadadze and
Porrati. In the model the present accelerating expansion of the universe 
is attributed to leaking gravity into extra dimension
\cite{Dvali:2000hr}.
This idea reproduces the present cosmic acceleration 
without dark energy component,
and consequently predicts the modification of Newton's law at
cosmological scales.
Another interesting proposal is a braneworld model with
Gauss-Bonnet term, which  suggests 
$G_{\rm N}=\frac{2+ \bar{\alpha}}{3\bar{\alpha}}G_{\infty}$ with 
a model parameter $\bar{\alpha}$ \cite{Deruelle:2003tz}.

These theoretical suggestions indicate that it is quite
important to place possible constraints on the value of $G$ at
astronomical and cosmological scales. 
In this paper, we take a simple parameterization of 
\begin{equation}
    G(r)= \xi G_{\rm N}~,
\end{equation}
for a finite range of $r$ relevant to the measurement 
which we are considering. As such a measurement, we consider
CMB anisotropies and primordial abundances created at BBN
epoch, and put the constraints on the value of $\xi$ 
at two different scales relevant to these observations. 
The cosmological constraints are compared with 
an astronomical constraint determined from the isochrone of
globular clusters.


The observed primordial light-element abundances constrain the value of $G$ 
during the BBN epoch 
from the time of weak reaction
freezeout ($t \sim 1 ~\mathrm{sec}, ~T \sim 1 ~\mathrm{MeV}$) to the
freezeout of nuclear reactions ($t \sim 10^4 ~\mathrm{sec}, ~T \sim 10
~\mathrm{keV}$). In this epoch, the length of cosmic horizon varies from
$10^8 ~\mathrm{m}$ to $10^{12} ~\mathrm{m}$, and thus BBN can constrain 
Newton's constant at these scales.

The primordial helium abundance is obtained by
measuring extra galactic HII regions. We adopt range of $\yp = 0.2452
\pm 0.0015$ \cite{Izatov:1999} for the helium abundance. 
The primordial deuterium
is best determined from its absorption lines in high redshift Lyman
$\alpha$ clouds along the lines of sight to background quasars. For
deuterium there is a similar possibility for either a high or low
value. For the present discussion, however, we shall adopt the generally
accepted low value for the $\dovh$ abundance, $\dovh
=2.78^{+0.44}_{-0.38} \times 10^{-5}$ \cite{Kirkman:2003}.

The increase of Newton's constant causes the increase of the
universal expansion rate. This makes the neutron-to-proton ratio 
larger, because the weak reactions freezeout at a higher temperature, 
and also because there is less time for neutrons to decay 
between the time of weak-reaction freezeout and the onset of BBN. 
Consequently, larger 
value of $G$ during BBN epoch yields a larger $\he4$ abundance, 
since most of the free neutrons are converted into $\he4$ nuclei. 
$\dovh$ also
increases largely because the reactions destroying deuterium fall out of
nuclear statistical equilibrium while the deuterium abundance is
higher. Similarly, there is less time for the destructive reaction $\li7
(p,\alpha)\he4$. This causes $\li7$ to be more abundant for 
$\eta < 3 \times 10^{-10}$. However, there is also less time for the $\he4
(\he3,\gamma)\be7$ reaction to occur. This causes $\li7$ to be less
abundant for $\eta > 3 \times 10^{-10}$ \cite{Ichiki:2002eh}. 

The upper limit of Newton's constant comes from  $\he4$
upper bound and $\dovh$ upper bound. The lower limit comes from the
lower bounds. 
We note that the constraint from $\li7$  is not consistent with
those from $\he4$ and $\dovh$, even when we vary $\xi$. In the present 
analysis, however, we omit the constraint from $\li7$ abundance, since
it involves an uncertainty more largely than the other primordial elements do. 
The BBN constraint thus obtained is $0.95 \le \xi \le 1.01$.

\begin{figure}
\rotatebox{-90}{\includegraphics[width=6cm]{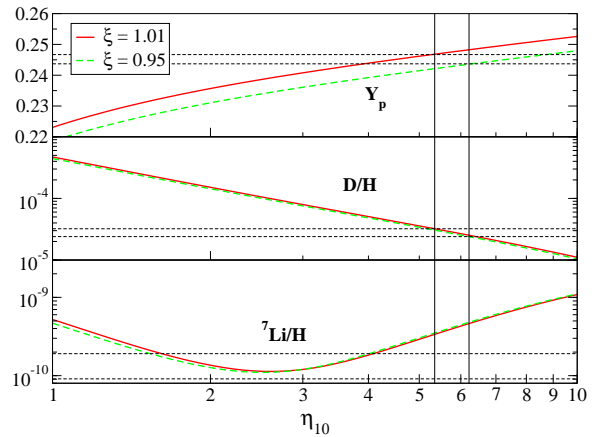}}
\caption{Predicted BBN light-element abundances v.s. the baryon-to-photon
 ratio $\eta_{10}$.  $\he4$, D and $\li7$ abundances are shown in the
 top, center and bottom panels, respectively. They are compared with
 the observationally inferred primordial abundances
 (horizontal lines).
 Plotted are models with $\xi=1.01$ (solid) and
 $\xi=0.95$ (dashed). In our analysis neutron life time is taken to be 
 the average value of $\tau_n = 878.5$ sec \cite{Serebrov:2004} and
 $\tau_n = 885.7$ sec \cite{Particle:2004}.}
\end{figure}
\begin{figure}
\rotatebox{-90}{\includegraphics[width=6cm]{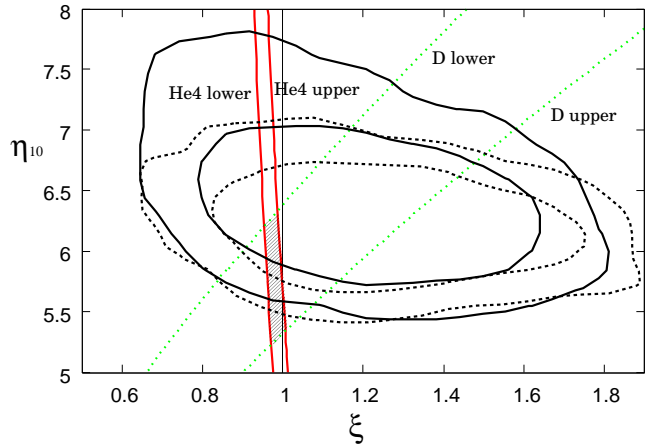}}
\caption{Constraints from the primordial abundances (lines) and CMB (contours)
on $\eta_{10}$ and $\xi$. 
 The shaded region denotes parameters allowed by BBN. The contours show the
 marginalized $1,2 \sigma$ limits on this parameter plane from fits to
 the CMB power spectrum.
The solid and dashed contours correspond to
 the limits from WMAP data alone, and  those from WMAP, CBI and ACBAR data
 sets, respectively.}  
\end{figure}


The temperature fluctuations at recombination observed through the
CMB anisotropies contain much information on many kinds of cosmological
parameters and evolution of perturbations at a wide range of scales.
Typically, the scale which can be explored by CMB observations
currently available is from the horizon scale at present ($\sim$ Gpc)
to $\sim 10$ Mpc in comoving coordinate. 
This shows that the scales relevant to CMB are from $10^2$ pc to $\sim$
Gpc, 
if we consider the evolution of perturbations from horizon crossing of
each Fourier mode;
for example, $10^2$ pc is the horizon scale at the time when
the mode of $\sim 10$ Mpc in comoving coordinate enters the
horizon.
Thus it follows that CMB can constrain the value of $\xi$ 
at scales larger than $\sim 10^2$ pc.
Here, in order to calculate CMB anisotropies in a consistent manner we
assume that the scale dependence of Newton's constant is very weak at 
the relevant scales, which is consistent with a simple parameterization
of Eq. (1).

In order to obtain a constraint on $\xi$ from latest CMB anisotropy
data sets, we generate CMB angular power spectra $C_\ell$ in a
wide range of $\xi$ by using a Boltzmann code of CMBFAST
\cite{Seljak:1996is}.
It is well known, however,  that in addition to  
$\xi$ there exist many other cosmological parameters relevant to CMB. 
Thus, we explore the likelihood in seven dimensional parameter space,
i.e.,  $\Omega_{\rm b} h^2$ (baryon density), $\Omega_{\rm c} h^2$ (cold
dark matter density), $h$ (Hubble parameter), $z_{\rm re}$ (reionization
redshift), $n_{\rm s}$ (power spectrum index), $A_{\rm s}$ (overall
amplitude), and $\xi$. We then marginalize over nuisance parameters 
through the use of Markov Chain Monte Carlo technique \cite{Lewis:2002}.

\begin{figure}
\rotatebox{-90}{\includegraphics[width=6cm]{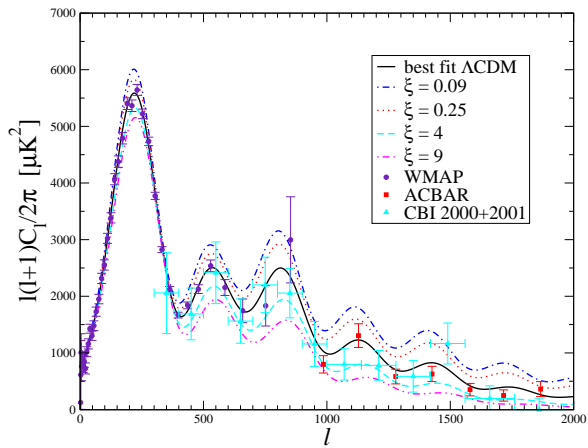}}
\caption{CMB power spectrum with and without the variation of
 $\xi$. {Higher peaks are more severely damped as $\xi$ increases, 
 while the height of the first peak is almost unchanged.}}
\label{fig3}
\end{figure}

The most distinguishable effects of changing Newton's constant 
appear at the amplitude of the acoustic peaks in
the CMB power spectrum as shown in Fig. \ref{fig3}.
The main reason for this is that, as already found in
\cite{Zahn:2003}, the visibility function,
 $g(\tau)=\dot{\kappa} \exp(-\kappa)$,
changes with $\xi$, where $\kappa$ is the optical depth of the Thomson
scattering.
More specifically, increasing Newton's constant makes the
expansion of the universe faster at a give redshift, and makes it
more difficult for proton and electron to recombine to form hydrogen atom.
This leads to larger ionization fraction and broader visibility
function at last scattering epochs, which damp the anisotropies at small
scales due to canceling effect.

Second, in addition to the effect discussed above, we find
that increasing Newton's constant suppresses the second and
higher acoustic peaks even larger since for the increase of $\xi$ 
the scale of the first
acoustic peak ($\sim t_{\rm dec} \propto \xi^{-1/2}$) 
is shifted more largely than the diffusion scale for photons
to spread through the random walk does, ($\sim t_{\rm dec}^{1/2} \propto
\xi^{-1/4}$). 
Thus, the shape of acoustic peaks can be used to constrain the
variation of $\xi$.

\begin{figure}
\rotatebox{-90}{\includegraphics[width=6cm]{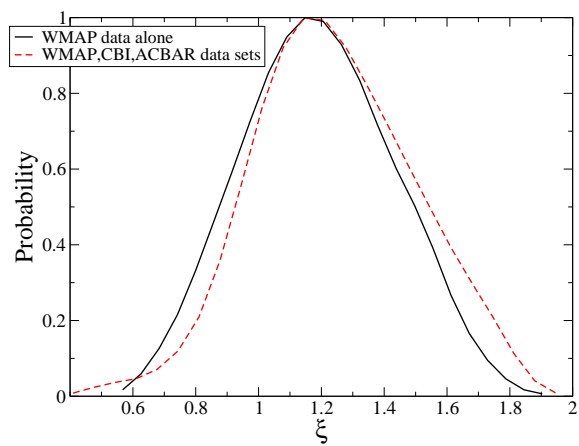}}
\caption{Marginalized probability distribution of $\xi$. The solid and
 dashed lines correspond to the probability distributions obtained by
 WMAP data alone and by WMAP, CBI and ACBAR data sets,
 respectively.}
\label{fig4}
\end{figure}

Figure \ref{fig4} shows the marginalized likelihood of $\xi$. 
We obtain from the figure that
$0.74 \la \xi \la 1.66$ by WMAP data alone \cite{wmap}, 
$0.75 \la \xi \la 1.74$ by WMAP, CBI and ACBAR data sets 
\cite{cbi3,acbar}, at $95\%$ confidence level.

Another constraint on the value of Newton's constant
can be obtained by analyzing the age of stars in globular clusters.
The key idea is that increasing Newton's constant causes
stars to burn faster \cite{Teller:1948}.
Thus, this allows us to constrain $\xi$ at stellar scale 
$\sim 10^9$ m, as we shall see below, 
by analyzing the timing of main sequence turn off.

Let us assume that luminosity of star depends on Newton's constant $G$
and helium abundance $Y$, approximately as $L \propto y(Y)g(G)$, where
$y$ and $g$ are functions of $Y$ and $G$
\cite{Degl'Innocenti:1995nt}. Since helium production 
should be proportional to the luminosity, we have $\frac{dY}{dt}\propto
y(Y)g(G)$. A star which departs from the main sequence today ($t_0$)
should be considered to have $Y\approx 1$ at its center so that
$\int^1_{Y_{\rm init}} \frac{dY}{y(Y)}\propto g(G) \int^{t_0}_{t_{\rm
init}} dt$. We further assume that $g(G) \propto G^\gamma$, where
$\gamma \approx 5.6$ have been obtained from numerical simulation
\cite{Degl'Innocenti:1995nt}. From the fact that the l.h.s. of the
above equation does not depend on $G$ and time, we have the relation,
\begin{equation}
 \tau_* = \xi^{\gamma}\int^{t_0}_{t_0-\tau}dt=\tau
 \xi^\gamma~.
\end{equation}
Here $\tau_*$ is the apparent turn-off age, which should
be obtained by analyzing HR diagram of a globular cluster with the
standard value of $G$, and $\tau$ is the true age of the globular
cluster.
Thus, if information on the true age of globular cluster is available,
the globular cluster can be used to constrain $\xi$;
\begin{equation}
 \left(\frac{\tau_{\rm max}}{\tau_*}\right)^{-{1 \over \gamma}} \la \xi
   \la   \left(\frac{\tau_{\rm min}}{\tau_*}\right)^{-{1
   \over \gamma}} ~.
\end{equation}
If we take $\tau_{\rm max}=15.8$ (2 $\sigma$ upper bound on the
expansion age of the universe obtained by our analysis in Sec. III
including the variation of $\xi$), and
conservatively assume that $\tau_{\rm min}=10$ Gyr
\cite{Degl'Innocenti:1995nt}, we then obtain
\begin{equation}
 0.93 \la \xi \la 1.09~,
\end{equation}
where we use $\gamma = 5.6$ and $\tau_* = 12.9\pm2.9$ Gyr, which is age of
the galactic globular clusters \cite{Carretta:2000}.

\begin{figure}[h]
\rotatebox{-90}{\includegraphics[width=6cm]{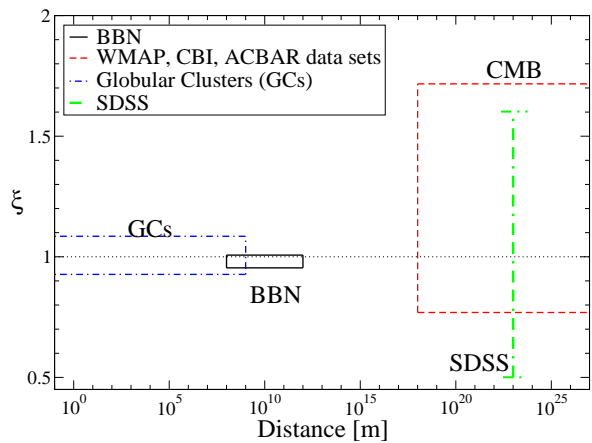}}
\caption{Limits on the variation of $\xi$
 from various observations. Limit from SDSS is taken from
 \cite{Shirata:2005yr}.}
\label{fig5}
\end{figure}

All the higher-dimensional theories of gravity proposed recently
allow Newton's constant to be scale-dependent. In this paper, assuming the 
dependence is weak for horizon scales in BBN epoch 
($10^8 \sim 10^{12}~$m) and also for those in CMB epoch ($10^2 \sim
10^9~$ pc), 
we place constraints on $\xi=G/G_{\rm N}$ at the cosmological scales. 
An important point is that the present analysis yields constraints on 
the value of $G$ itself, while other astronomical 
tests of the inverse
square law do so only on the value of $GM$ including unknown 
mass $M$ of interacting bodies.

Increasing Newton's constant enhances the universal expansion rate,
and then leads to larger helium and deuterium abundances produced at BBN
epoch. 
We have re-examined this effect including the latest experimental data
on the neutron lifetime \cite{Serebrov:2004,Mathews:2004kc}. We found that
the experimental value $G_{\rm N}$ ($\xi=1$) is now quite consistent with the
observed abundances of primordial light elements, and the variation of
Newton's constant is tightly constrained to $0.95 \la \xi \la 1.01$.

The variation of Newton's constant also affects the power spectrum of
CMB anisotropies through the change of the recombination and 
photon diffusion processes. 
We found that the difference emerges at smaller scales. Thus, 
observations at higher multipoles are essential to put 
a tighter constraint. However, even when higher multipoles data currently
available from CBI and ACBAR are included,
we found no improvement in constraint on $\xi$, because of scatters
in data at higher multipoles. WMAP data alone place a constraint: $0.74
\la \xi \la 1.66$. If we combine CBI and ACBAR data sets, the constraint
becomes  $0.75 \la \xi \la 1.74$.

In Fig. \ref{fig5}, we summarize results of the current work.
The value of $\xi$ is fixed to one at laboratory scale $\sim 1~$m by
direct experiments. We have two possibilities of transition from short
distance regime where $G=G_N$ to long distance one where $G=\xi G_N$; one is geophysical scale (i.e., $\sim 1$ km - $100$ km,
where the constraints on the inverse square law are relatively
weak), the other is scale 
beyond the solar system ($\ga 10^{13}$ m), 
where we have only poor knowledge on $G$. 
If we consider the former case, 
BBN gives the tightest constraint on $\xi$.  
Globular cluster also gives
a consistent but weaker constraint. On the other hand, if we consider
the latter case, CMB anisotropies and galaxy clustering
\cite{Sealfon:2004gz,Shirata:2005yr} are the only observations to put constraints on 
$\xi$. Thus, higher precision measurements of CMB anisotropies,
particularly in its higher multipoles are highly expected 
to  determine the value of $G$ at large scales beyond the solar system 
and then to confirm the 
necessity of the higher-dimensional theories of gravity.

{\it Acknowledgments:---}
One of the authors (K.I.) would like to thank T. Chiba for helpful
discussions. K.I.'s work is supported by Grant-in-Aid for JSPS fellows. 
This work is supported in part by Grant-in-Aid for Scientific
Research(14540271).

\end{document}